\documentclass[twocolumn,aps,prl,showpacs,floatfix,,superscriptaddress]{revtex4}
\usepackage[latin9]{inputenc}
\usepackage{graphicx}
\usepackage{esint}

\makeatletter
\@ifundefined{textcolor}{}
{%
 \definecolor{BLACK}{gray}{0}
 \definecolor{WHITE}{gray}{1}
 \definecolor{RED}{rgb}{1,0,0}
 \definecolor{GREEN}{rgb}{0,1,0}
 \definecolor{BLUE}{rgb}{0,0,1}
 \definecolor{CYAN}{cmyk}{1,0,0,0}
 \definecolor{MAGENTA}{cmyk}{0,1,0,0}
 \definecolor{YELLOW}{cmyk}{0,0,1,0}
}

\usepackage{bm}\usepackage{float}\usepackage{afterpage}

\makeatother

\begin{document}

\title{Complete phase diagram and topological properties of interacting bosons in one-dimensional superlattices}

\author{Tianhe Li}
\affiliation{Department of Physics, Beihang University, Beijing, 100191, China}

\author{Huaiming Guo}
\thanks{hmguo@buaa.edu.cn}
\affiliation{Department of Physics, Beihang University, Beijing, 100191, China}
\affiliation{Department of Physics, The University of Hong Kong, Pokfulam Road,
Hong Kong}

\author{Shu Chen}
\affiliation{Beijing National
Laboratory for Condensed Matter Physics, Institute of Physics,
Chinese Academy of Sciences, Beijing 100190, China}
\affiliation{Collaborative Innovation Center of Quantum Matter, Beijing, China}

\author{Shun-Qing Shen}
\affiliation{Department of Physics, The University of Hong Kong, Pokfulam Road,
Hong Kong}

\begin{abstract}
The interacting bosons in one-dimensional inversion-symmetric superlattices are investigated from the topological aspect. The complete phase diagram is obtained by an atomic-limit analysis and quantum Monte Carlo simulations and comprises three kinds of phases: superfluid, persisted charge-density-wave and Mott insulators, and emergent insulators in the presence of nearest-neighbor hoppings. We find that all emergent insulators are topological, which are characterized by the Berry phase $\pi$ and a pair of degenerate in-gap boundary states. The mechanism of the topological bosonic insulators is qualitatively discussed and the ones with higher fillings can be understood as a $\frac{1}{3}$-filling topological phase on a background of trivial charge-density-wave or Mott insulators.
\end{abstract}

\pacs{ 03.65.Vf, 
  73.21.Cd 
 37.10.Jk 
 67.85.Hj 
 }

\maketitle
\section{Introduction}
Topological phases of matter have bulk gap but gapless boundary states \cite{rev1,rev2,rev3,rev4}. These studies have been extended to interacting fermions and bosons. Specially much progress has been made for one-dimensional (1D) topological bosonic phases. Classification schemes have been proposed for such phases \cite{bcla0,bcla1,bcla2}. It is recognized that long-known 1D Haldane phase for spin-1 chain is a topological bosonic phase \cite{haldane}.
While these studies are important for our understanding of topological bosonic states, it is also highly desirable to study nontrivial bosonic phases in experimental accessible systems.

Motivated by the experimental progress in studying topological phases with ultracold atoms \cite{cold1}, exploring topologically nontrivial phases of interacting bosons in 1D superlattices has attracted intensive attentions.
It is found that by simply replacing free fermions with interacting bosons in 1D topological superlattices \cite{tf1,tf2,su1}, the resulting extended Bose-Hubbard models display nontrivial topological property \cite{tb1,tb2,tb3,tb4,tb5}, which are well understood from the hard-core limit. As the Bose-Hubbard model exhibits rich quantum phases due to multiple occupations of bosons on a single site \cite{bhm1,bhm2,jordan}, one may expect some novel phenomena emerging in topologically nontrivial superlattice systems beyond the hard-core limit. As we shall display in this work, surprisingly plentiful phase diagram, including the emergence of topological bosonic insulators at both fractional and integer fillings and various phase transitions induced by varying the chemical potential and the strength of superlattice potential, are found in the simple superlattice system, which is realizable in cold-atom experiments \cite{cold2,Roati}.

We firstly determine the complete phase diagram of the system by a combination of atomic-limit analysis and quantum Monte Carlo (QMC) simulations.
While the phase diagram in the atomic limit contains charge-density-wave (CDW) and Mott insulating phases with various fillings, these insulating phases persist but are separated by superfluid phases when hopping terms are included.
It is interesting that some nontrivial insulating phases emerge between two adjacent persisted insulators and these emergent insulators are topologically nontrivial, characterized by the Berry phase $\pi$. On the other hand,  the persisted insulators are topologically trivial as they are adiabatically connected to the insulators in the atomic limit.
The nontrivial topological property of the emergent insulators is further confirmed by the presence of a pair of degenerate in-gap boundary states under open boundary condition (OBC), which leads to the splitting of topological plateau in the $\mu-\rho$ curve. By varying the superlattice strength, one may observe phase transitions between the topologically different insulating phases.
Our results present a complete understanding on the topological phase diagram of interacting bosons in 1D superlattices,
which may shed light on the experimental exploration of the predicted exotic topological phases and phase transitions.

\section{Model of interacting bosons in 1D superlattices}
We consider the interacting bosons loaded into the optical superlattice with inversion symmetry in the grand canonical ensemble, whose basic physics is described by the extended Bose-Hubbard model with a superlattice potential:
\begin{eqnarray}\label{eq1}
\hat{H}=-t\sum_j (\hat{b}_j^{\dagger}\hat{b}_{j+1}+h.c.)+\sum_j V_j \hat{n}_j   \\ \nonumber
+U\sum_j \hat{n}_j(\hat{n}_j-1)-\mu \sum_j \hat{n}_j,
\end{eqnarray}
where $\hat{b}_j$ ($\hat{b}_j^{\dagger}$) is the bosonic annihilation (creation) operator, $\hat{n}_j=\hat{b}_j^{\dagger}\hat{b}_j$ is the number operator of bosons, the hopping amplitude $t$ is set to be the unit of the energy $(t=1)$, $V_{j+T}=V_{j}$ represents a superlattice potential with the period $T$,
$U$ represents the strength of on-site interactions, and $\mu$ is the chemical potential. The inversion symmetry further requires $V_j=V_{T+1-j}$ with $j=1,...,T$, which can be realized by a simple bichromatic superlattice potential $V_j=A\cos(2\pi j/T+\delta)$ by tuning the phase to $\delta_{1,2}=\pi(1-1/T)$ or $\pi(2-1/T)$, with $A$ the strength of the potential \cite{Roati}. Without loss of generality, we shall focus our study on the case with $T=3$ and $\delta=2\pi/3$ in the following calculations.

 In the hard-core limit $U=\infty$, Eq.(\ref{eq1}) can be mapped to the free fermionic model in the superlattice \cite{jordan}, which has been shown that 1D topological phases protected by inversion symmetry exist at inversion-symmetric point $\delta_{1,2}$ for fillings of $1/T$, $(T-1)/T$ respectively \cite{spt}. The 1D topological phase is characterized by the Berry phase, $\gamma=\oint {\cal A}(k) dk$, with the Berry connection ${\cal A}(k)=i\langle u_k|\frac{d}{dk}|u_k\rangle$ and $u_k$ the occupied Bloch state \cite{berry1,berry2,zak}. Due to the protection of the inversion symmetry, the Berry phase $\gamma$ mod $2$ takes two values: $\pi$ for a topological phase and $0$ for a trivial phase. Corresponding to the nontrivial Berry phase $\pi$, there appear a pair of degenerate in-gap states under OBC, whose distributions are localized near the boundaries.

\section{Phase diagram of softcore case}
\begin{figure}[htbp]
\centering \includegraphics[width=7cm]{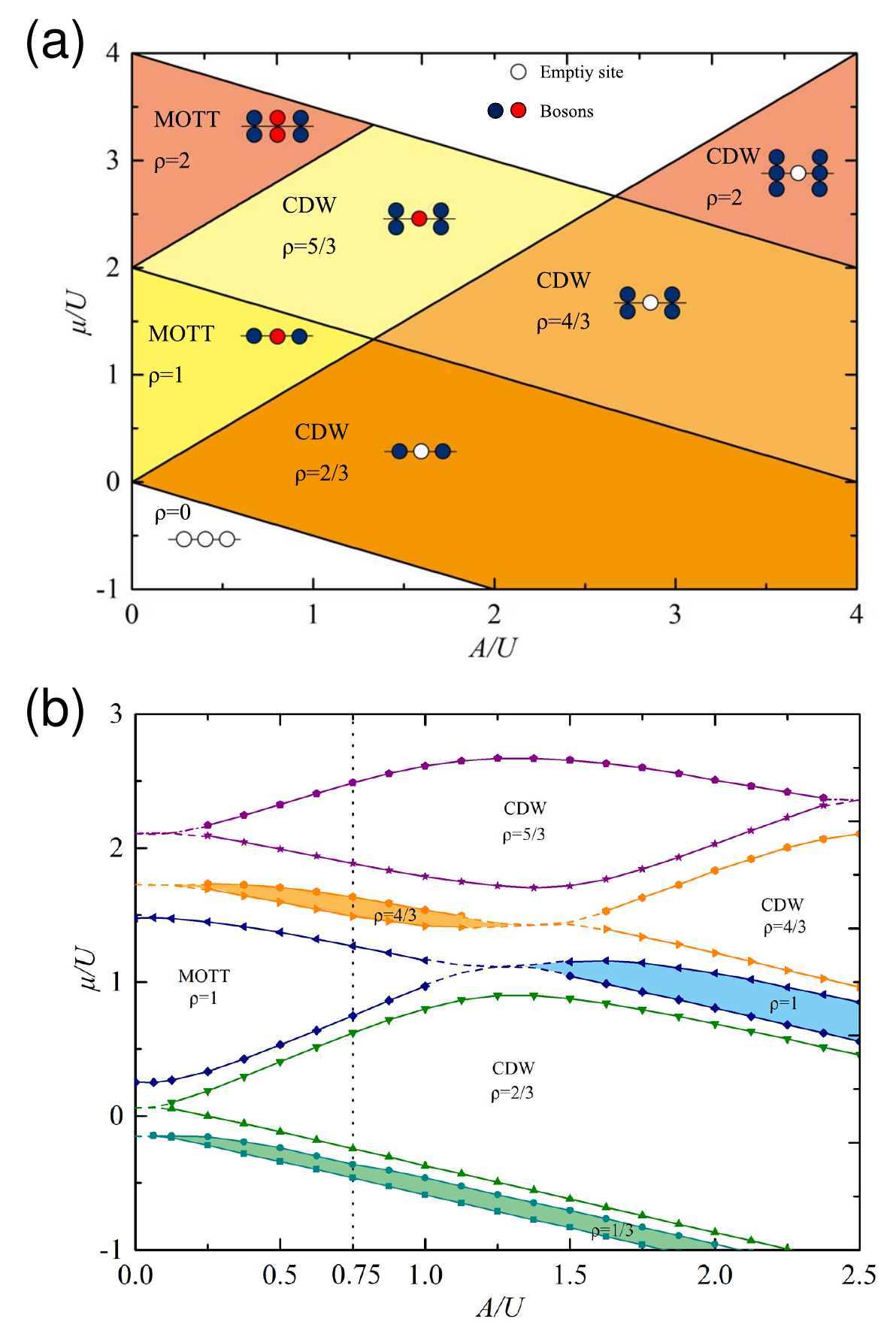} \caption{(Color online) (a) The phase diagram of softcore bosons in the atomic limit. (b) The phase diagram of softcore bosons in the presence of NN hoppings, in which the shaded colored regions represent topological phases and the dotted vertical line is a typical cut along which the detail of the simulation will be shown in next figure. The phase of the superlattice potential is $\delta=\frac{2\pi}{3}$, when the middle (side) site of the unit cell has high (low) potential energy.}
\label{fig1}
\end{figure}

It is useful to firstly consider the atomic limit $t=0$. Whether a boson can be added to the $j$th site with $n_j$ bosons is determined by the energy difference $\Delta E=E(n_j+1)-E(n_j)=-\mu+V_j+2Un_j$ with $E(n_j)=-\mu n_j+V_j n_j+Un_j(n_j-1)$ the total energy of the bosons on the $j$th site. If $\Delta E<0$, the total energy is lowered and one more boson can be added to the site. Thus a series of lines determined by $-\mu+V_j+2Un_j=0$ separate different insulating phases. The phase diagram of the atomic limit in the $(A/U, \mu/U)$ plane is shown in Fig.\ref{fig1}(a). The set of parallel lines $\mu/U=A/U+2n_j$ determine the occupation on the middle site of the unit cell and the separated regions have gradually increasing occupation numbers. The set of parallel lines $\mu/U=-A/2U+2n_j$ determine the occupation on the side sites. There are two kinds of insulators in the phase diagram: Mott insulator with a uniform density $\rho=n_1$ (or $n_2$) with $n_1(n_2)$ the number of bosons on the side (middle) sites of the unit cell; CDW insulator with a density profile reflecting the modulation and an average density $\rho=\frac{2}{3}n_1+\frac{1}{3}n_2$. Different insulators are characterized by the values of structure factor $S(Q)$. For the Mott insulators, the structure factor has one peak at $Q=0$ with $S(Q)=\rho^2$. In the CDW insulator, a peak develops at $Q=\frac{2\pi}{3}$ with $S(Q)=\frac{(n_2-n_1)^2}{9}$ .

Next we turn on nearest-neighbor (NN) hoppings and the phase diagram obtained from QMC simulations is shown in Fig.{\ref{fig1}}(b). Compared to the atomic one, the phase diagram is considerably modified. Although each insulator in the atomic limit persists, the phase boundaries are deformed and incommensurate superfluid regions appear between the commensurate insulating regions. Moreover between two adjacent regions separated by the line $\mu/U=-A/2U+2n_j$, a insulator with intermediate filling $\rho=(\rho_1+\rho_2)/2$ emerges ( $\rho_1, \rho_2$ are the fillings of the two regions). Except of the insulator with $\rho=1/3$, the emergent insulating regions are connected to the persisted ones with the same fillings, e.g., the shaded regions in Fig.{\ref{fig1}}(b) with $\rho=1$ and $\rho=4/3$, and are separated from others by superfluid phases.

\begin{figure}[htbp]
\centering \includegraphics[width=7.5cm]{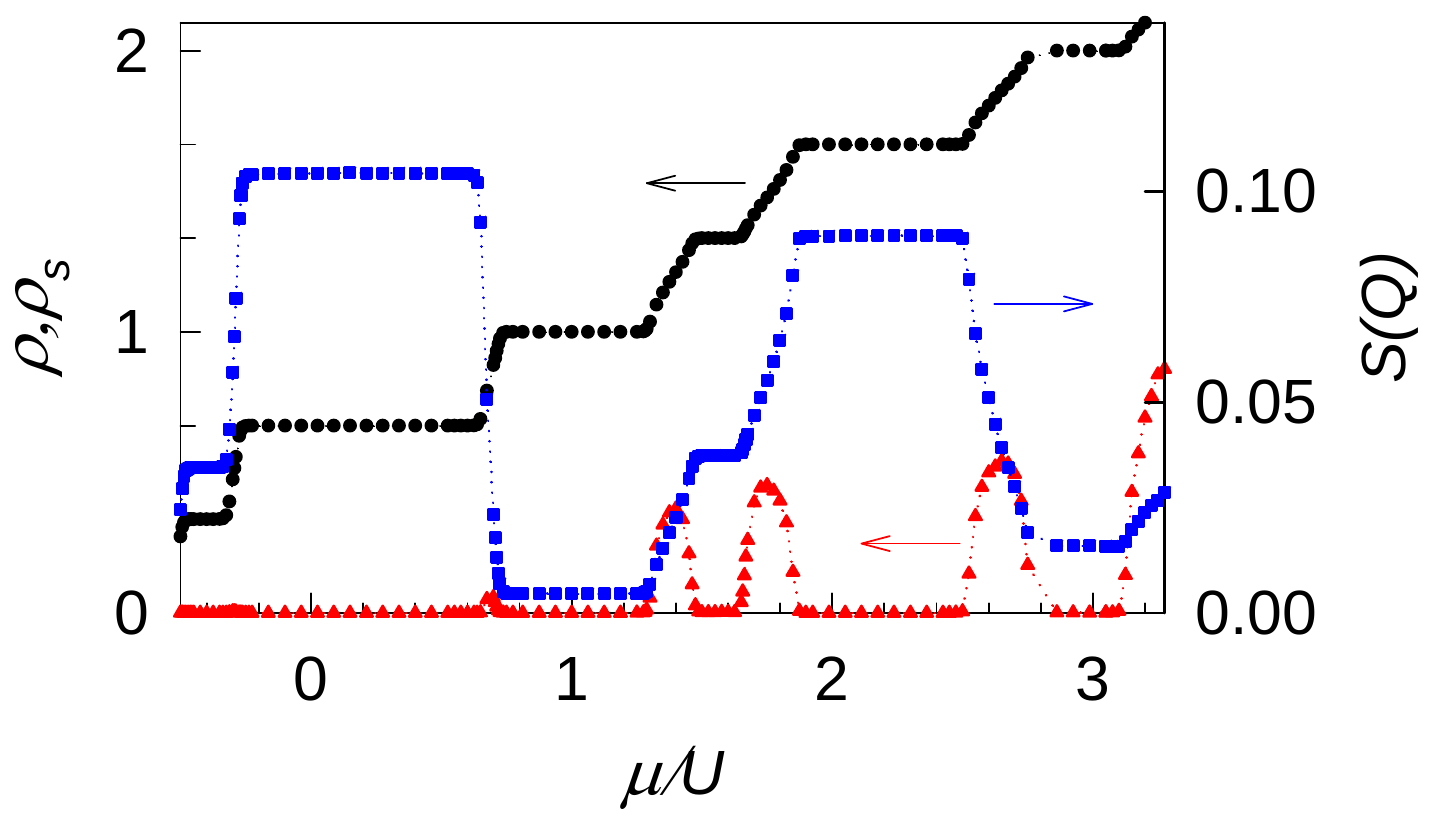} \caption{(Color online) The average density, the superfluid density and the structure factor as a function of $A$ along the cut with fixed $A/U=0.75$ in the phase diagram.}
\label{fig2}
\end{figure}

The above phase diagram is obtained by calculating the structure factor $S(Q)$,
 \begin{eqnarray}\label{eq2}
 S(Q)&=&\frac{1}{L^2}\sum_{jk} e^{iQ(j-k)}\langle n_j n_k\rangle,
 \end{eqnarray}
 and the superfluid density $\rho_s=\frac{\langle W^2\rangle}{2\beta t}$ with $W$ the winding number and $\beta$ the inverse temperature \cite{sandvik}. An insulator is characterized by $S(Q)\neq 0$ and $\rho_s=0$, while a superfluid phase by $S(Q)=0$ and $\rho_s\neq 0$.
 To see the details, we present the numerical results on the cut with fixed $A/U=0.75$ in the phase diagram, along which all the typical insulating phases are met. As shown in Fig.\ref{fig2}, the average density $\rho$ exhibits various plateaus at commensurate fillings, on which the superfluid density $\rho_s$ vanishes. So the plateaus correspond to the imcompressive insulators, whose gaps are the widths of the plateaus. Between the insulators, the average density increases continuously with the chemical potential and the superfluid density is finite, implying the system is in superfluid phase. Among the insulators, those persisted Mott and CDW insulators are distinguished by the values of the structure factors (see the Appendix).

\section{Topological property of the insulators}
\begin{figure}[htbp]
\centering \includegraphics[width=8.5cm]{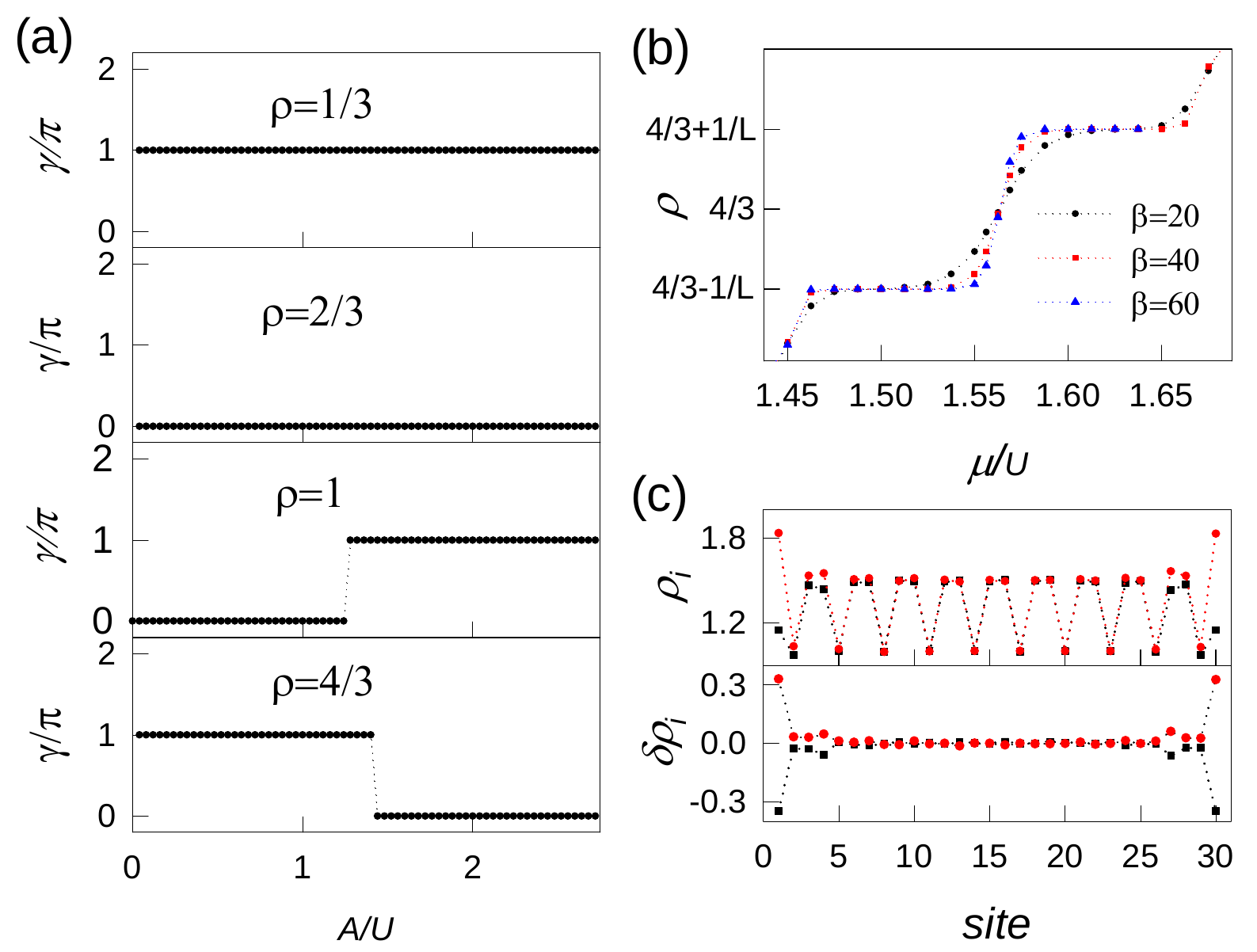} \caption{(Color online) (a) The Berry phase as a function of $A$ at several typical fillings. (b) The splits of the plateaus under OBC at the filling $\rho=\frac{4}{3}$. (c): (upper) The distribution of the bosons on two representative points of the split plateaus in (b): black square ($\mu=1.5$), red circle ($\mu=1.625$); (lower) the difference $\delta \rho_i$ compared to the bulk sites. The inverse temperature of (c) is $\beta=60$. In (b) and (c) the parameter $A/U=0.75$ is used.}
\label{fig3}
\end{figure}

The topological property of the interacting bosonic insulators is characterized by the Berry phase defined with the twisted boundary phase $\theta$ \cite{exact,twist1},
 \begin{eqnarray}\label{eq3}
 \gamma=\oint i\langle \psi_{\theta}|\frac{d}{d\theta}|\psi_{\theta}\rangle,
 \end{eqnarray}
where $\theta$ takes values from $0$ to $2\pi$ and $\psi_{\theta}$ is the corresponding many-body ground-state wave-function. In Fig.\ref{fig3}(a), we calculate the Berry phase as a function of the strength of the superlattice potential $A$ at different fillings. The $\rho=\frac{1}{3}$ ($\rho=\frac{2}{3}$) insulators are in one piece of the phase diagram. For the $\rho=\frac{1}{3}$ insulators, the Berry phase is $\pi$ and they are topological. However for the $\rho=\frac{2}{3}$ insulators, the Berry phase is $0$, thus they are trivial. There are two pieces for the insulators with $\rho=1$ ($\rho=\frac{3}{4}$), which are connected at a critical point. For $\rho=1$ insulators, the Berry phase changes its value from $0$ to $\pi$ at the critical point, implying the $\rho=1$ insulators emerging at large $A$ are topological ones. However for $\rho=\frac{4}{3}$ insulators, the Berry phase changes its value from $\pi$ to $0$ at the critical point, and the $\rho=\frac{4}{3}$ insulators emerging at small $A$ are topological ones. The topological property of insulators at other fillings can be analyzed similarly. So phase transitions between topologically different insulators can be realized by tuning the superlattice strength. Moreover these results provide a complete understanding of the topological properties of the insulating phases exhibited by interacting bosons in 1D superlattices: all emergent insulators in the presence of NN hoppings are topological, while those persisted CDW or Mott insulators are trivial since they are adiabatically connected to the atomic ones.

Due to the bulk-boundary correspondence, there should appear a pair of degenerate in-gap boundary states under OBC in the topological phases. The plateau of the topological insulating phase in the $\mu-\rho$ curve should be altered: below the critical chemical potential corresponding to the energy of the in-gap states, none of the in-gap states are occupied and the average density changes to $\rho'=\rho-\frac{1}{L}$; above the critical chemical potential, both of the in-gap states are occupied and the average density changes to $\rho''=\rho+\frac{1}{L}$. It is verified by our QMC simulations. As shown in Fig.\ref{fig3}(b), the plateaus of the topological phases split into two pieces with a jump at the critical chemical potential and the magnitude of the jump is exactly $\frac{2}{L}$. Moreover the jump tends to be vertical in the limit of zero temperature, implying the two in-gap states are degenerate.
To verify the in-gap states are boundary ones, we calculate the distribution of the bosons under OBC on two representative points of the split plateaus. As shown in Fig.\ref{fig3} (c), the distribution on the bulk sites are nearly unchanged, thus the filling of the in-gap states happens near the boundaries. When none of the in-gap states are filled (the lower plateau), there is $\frac{1}{2}$- fractional boson less at each boundary compared to the bulk sites. However after both are filled (the higher plateau), there is $\frac{1}{2}$- fractional boson more. The results for the topological phases with other fillings are similar. As a contrast, we do not observe the splitting of plateaus and boundary states for trivial insulators. Thus our results give clear evidence that in the topological phases a pair of degenerate in-gap states appear and they are localized near the boundaries.

\begin{figure}[htbp]
\centering \includegraphics[width=7cm]{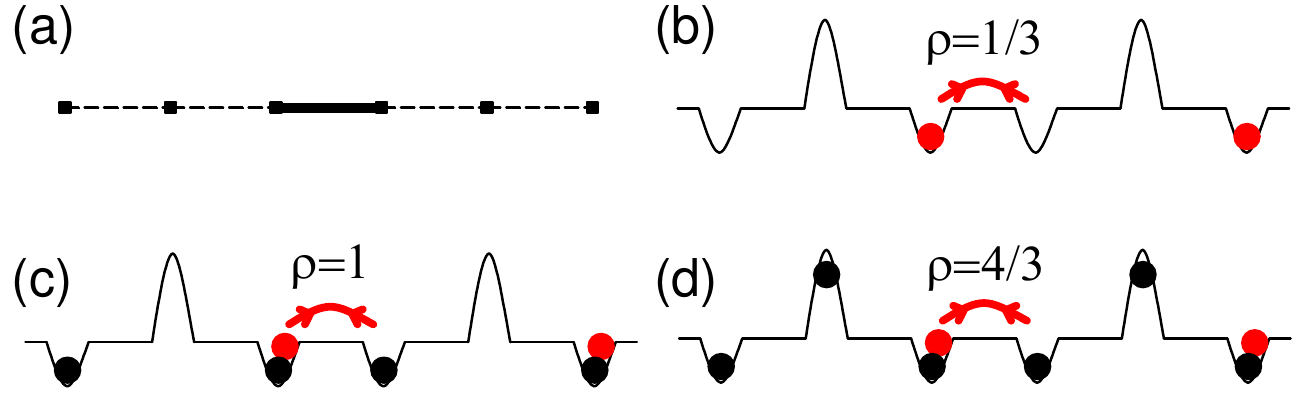} \caption{(Color online) (a) The topological trimerized model for comparison's purpose, in which the solid (dashed) line represents strong (weak) bond. Schematic illustration of the superlattice potential and the topological ground state at the filling: (b) $\rho=\frac{1}{3}$; (c) $\rho=1$; (d) $\rho=\frac{4}{3}$.}
\label{fig4}
\end{figure}

The nature of the topological phases can be qualitatively understood from the aspect of their analog to the famous topological trimerized model \cite{su2}, whose configuration is shown in Fig. \ref{fig4} (a). The main character of the model is that the bonds connecting different unit cells have larger hopping amplitude, while those inside a unit cell are equal and smaller. For the topological bosonic insulating phase at $\frac{1}{3}$ filling, the bosons tend to reside on the sites in the minima of the superlattice and averagely there is one boson on two adjacent low energy sites [see Fig.\ref{fig4} (b)]. Since the two sites are identical, the boson can hop freely between them with a larger amplitude to gain more kinetic energy. However the hoppings between other sites are barriered by the potentials and the amplitudes are relatively small. Thus the effective hopping amplitudes form a configuration similar to that of the topological trimerized model and the system exhibits nontrivial topological property. The topological phase at higher filling $\rho$ is similar except that it has a background of CDW or Mott insulator with $\rho=\rho_0-\frac{1}{3}$, e.g., the cases with the fillings $\rho=1, \frac{4}{3}$ shown in Fig.\ref{fig4} (c), (d). Furthermore the qualitative picture is consistent with the values of the structure factor (see the Appendix). So we can term the topological phases with higher fillings as topological bosonic CDW or Mott insulators depending on the background.

\begin{figure}[htbp]
\centering \includegraphics[width=7cm]{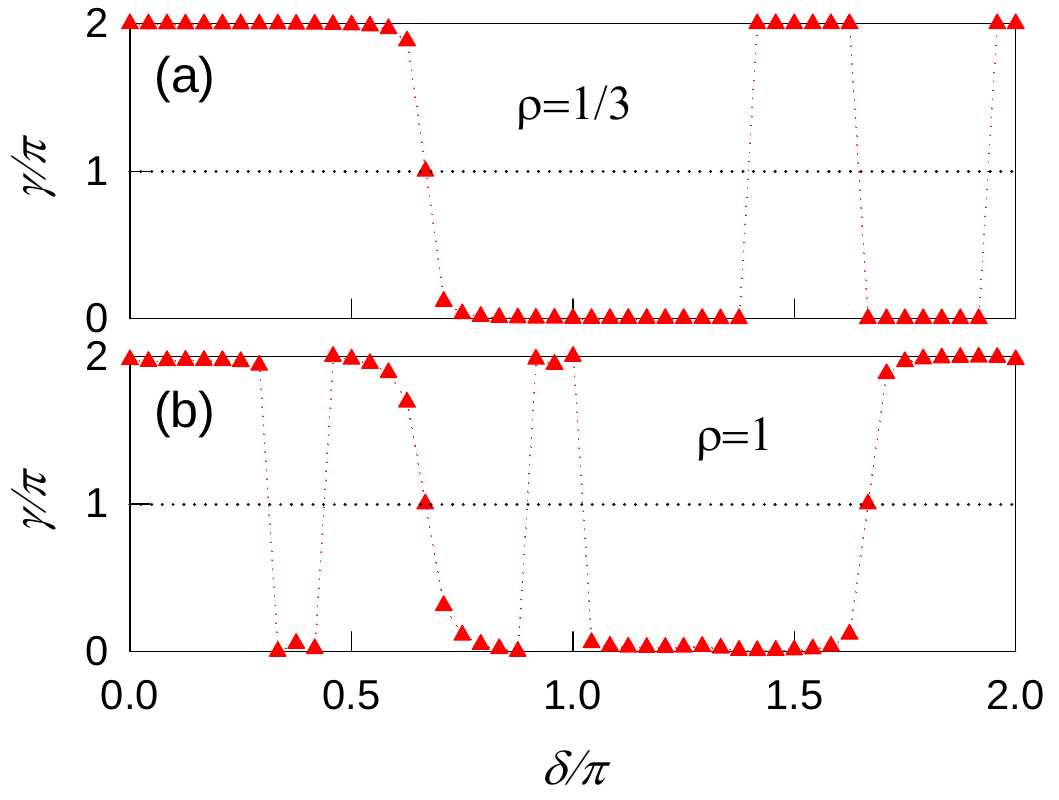} \caption{(Color online) The Berry phase as a function of the
phase $\delta$ of the superlattice at the filling: (a) $\rho = 1/3$ ; (b) $\rho = 1$.
Here the parameter $A/U = 2$ and the lattice size $L = 9$ are
used.}
\label{fig5}
\end{figure}

Finally we emphasize that although the system exhibits emergent insulating phases
at other $\delta$s where there are two identical sites in one unit cell, only those emerging at inversion-
symmetric points are topological. In Fig.\ref{fig5}, we calculate
the Berry phase as a function of $\delta$ at two typical fillings
for $A/U = 2$. It shows that the Berry phases of the insulators are only quantized to 0 or $\pi$ at inversion-symmetric
points. However when an insulator is emergent, its Berry phase
has nontrivial value $\pi$. So the identified topological bosonic insulators are protected by inversion symmetry, which provide concrete models for the classifications of 1D topological bosonic phases \cite{bcla0,bcla1}.

\section{Summary}
We study the topological phase diagram of interacting bosons in 1D superlattices with inversion symmetry. The complete phase diagram is obtained and the topological properties of the identified insulators are determined. It is found that the persisted CDW and Mott insulators are topological trivial since they are adiabatically connected to the atomic ones, while all emergent ones are topologically nontrivial. We present a qualitative mechanism for the topological bosonic insulators. This finding is of interests to cold-atom experiments. The studied model represents a simple experimentally accessible system and the various topological bosonic phases can be realized. One may use Bloch oscillations to measure the Berry phase \cite{cold2}, and {\it in situ} microscopy to detect the boundary states \cite{exp1,exp2,exp3}.

\begin{acknowledgments}
This work is supported by NSFC under Grants Nos. 11274032, 11104189 (H. G.); Nos. 11425419, 11374354 and 11174360 (S. C.), and the Research Grant Council of Hong Kong under Grant No. HKU7037 13P (S. S. and H. G.).
\end{acknowledgments}
\appendix
\section{More results from QMC simulations}
\begin{figure}[htbp]
\centering \includegraphics[width=7cm]{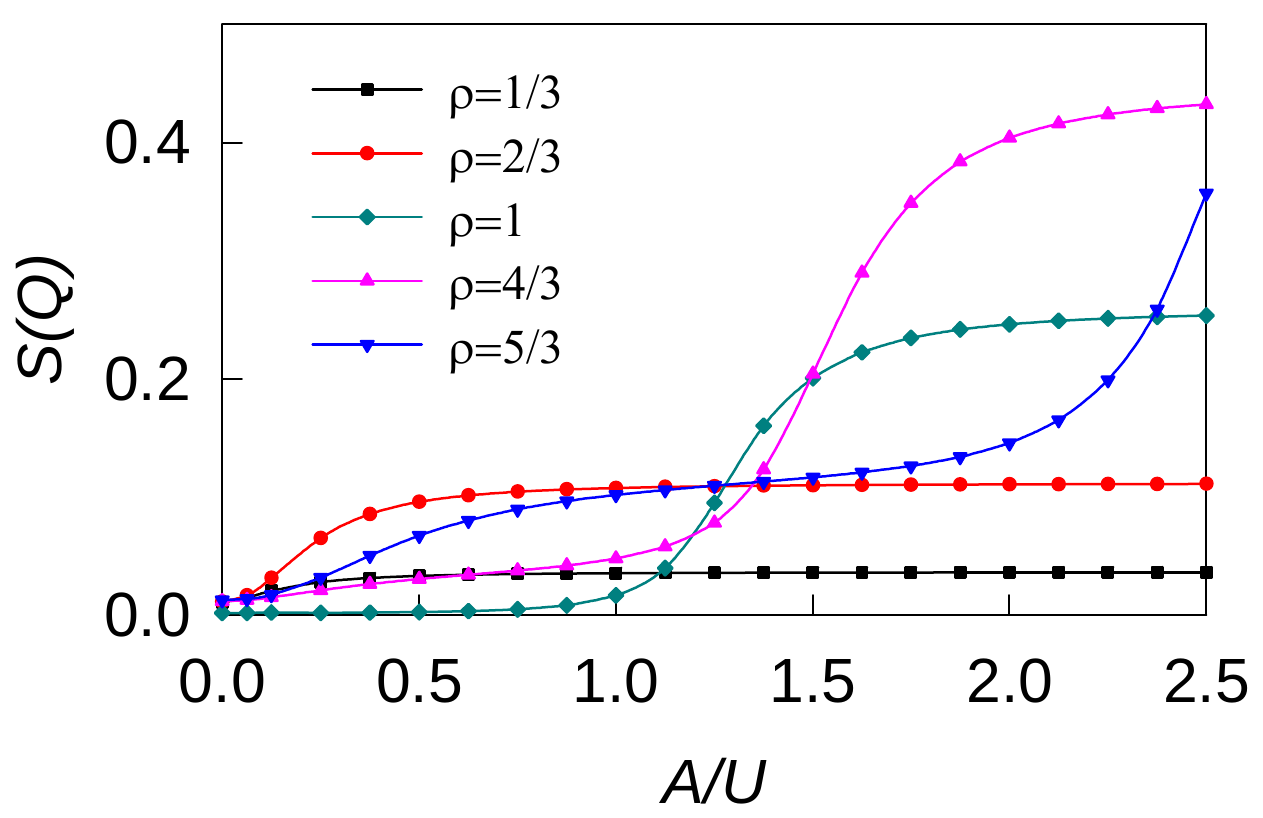} \caption{(Color online) The static structure factor $S(Q)$ at $Q=\frac{2\pi}{3}$ as a function of $A$ at different fillings.}
\label{fig6}
\end{figure}

We use QMC simulations to identify different insulators. Among them the trivial CDW and Mott insulators are distinguished by the values of the structure factor $S(Q)$ at $Q=\frac{2\pi}{3}$. However one should notice the fact that the limit values are reached when $A$ is away from the critical points, while the values deviate much from the atomic ones near the critical points, as shown in Fig.\ref{fig6}.

For the emergent nontrivial insulators, the values of the structure factor support our qualitative physical picture of the topological phases with higher fillings. In the picture, we can approximate: $n_j\approx n_j^{(0)}+\delta n_j$, with $n_j^{(0)}$ the boson number of CDW or Mott insulating backgrounds and $\delta n_j$ that of a $\frac{1}{3}$-filling topological phase. Then we have
$S(Q)\approx S_B(Q)+2S_{BT}(Q)+S_{T}(Q)$, where
\begin{eqnarray*}
S_B(Q)&=&\frac{1}{L^2}\sum_{jk} e^{iQ(j-k)}\langle n_j^{(0)} n_k^{(0)}\rangle, \\ \nonumber
S_{T}(Q)&=&\frac{1}{L^2}\sum_{jk} e^{iQ(j-k)}\langle \delta n_j \delta n_k\rangle, \\ \nonumber
S_{BT}(Q)&=&\frac{1}{L^2}\sum_{jk} e^{iQ(j-k)}\langle \delta n_j n_k^{(0)}\rangle. \nonumber
\end{eqnarray*}
The value of the structure factor in the topological phase away from the critical points can be approximated by the above formulas.

\begin{figure}[htbp]
\centering \includegraphics[width=7cm]{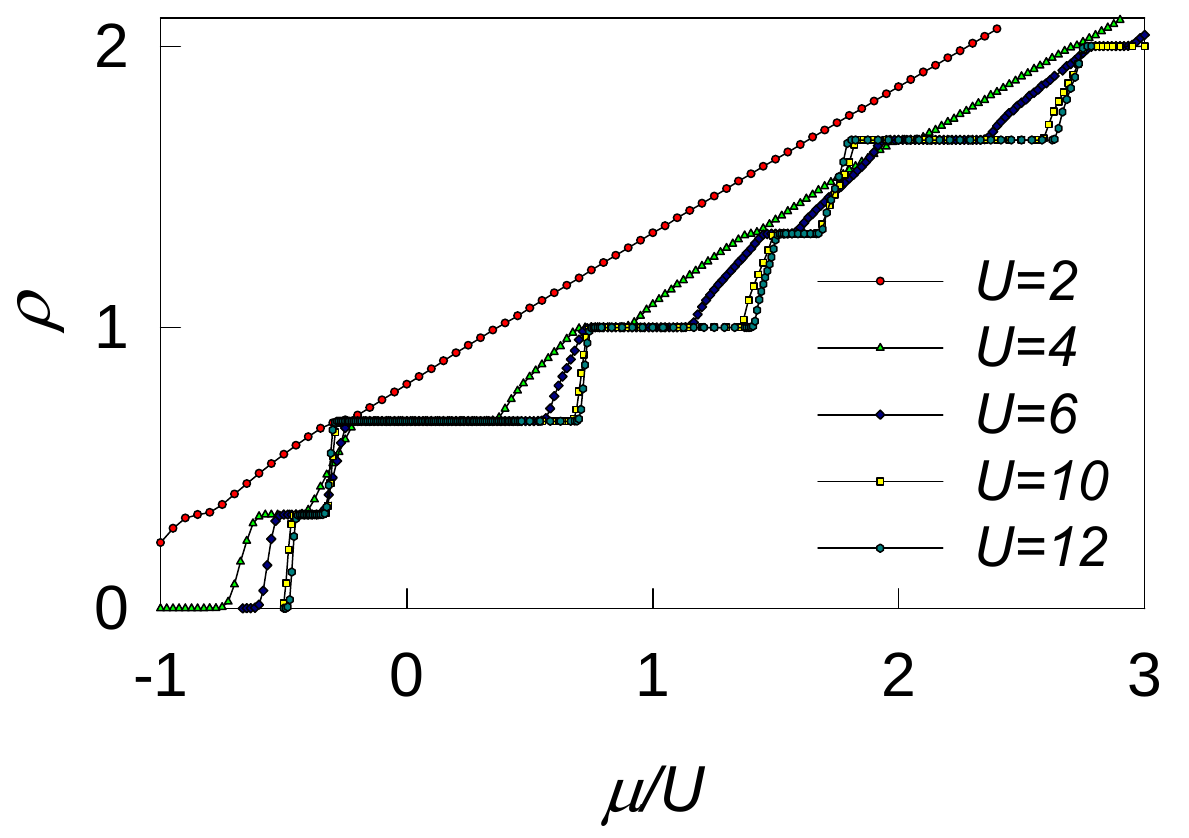} \caption{(Color online) The average density of bosons as a function of the chemical potential for different values of the on-site interaction $U$. Here $A/U=0.75$.}
\label{fig7}
\end{figure}

For the phase diagram of the softcore case, we take the on-site interaction $U=8t$. We also perform QMC simulations with other values of $U$. As shown in Fig.\ref{fig7}, some insulating phases tend to vanish when $U$ is decreased, which are replaced by the superfluid phases.

\begin{figure}[htbp]
\centering \includegraphics[width=8.5cm]{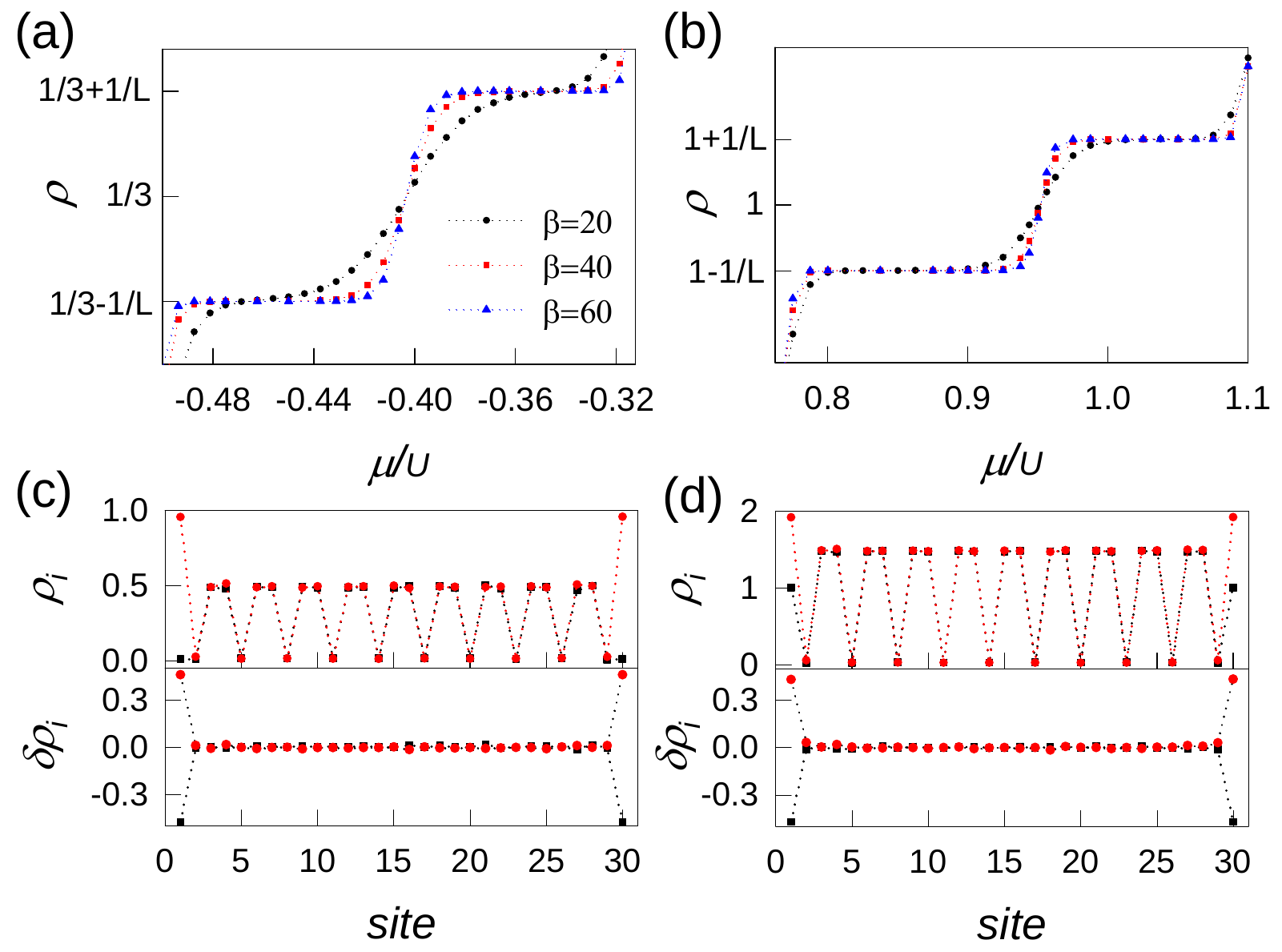} \caption{(Color online) The splits of the plateaus under OBC at the filling: (a) $\rho=\frac{1}{3}$; (b) $\rho=1$. The distribution of the bosons on two representative points of the split plateaus and the difference $\delta \rho_i$ compared to the bulk sites: (c) the filling $\rho=\frac{1}{3}$ and the chemical potential $\mu/U=-0.45, -0.35$; (d) the filling $\rho=1$ and the chemical potential $\mu/U=0.8625, 1.0375$. In (c) and (d) black square (red circle) represents the case with low (high) chemical potential.}
\label{fig8}
\end{figure}

We show the QMC simulations under OBC for the topological phase with the filling $\rho=\frac{4}{3}$ in Sec.IV. The splitting of topological plateau reflects the existence of the topological boundary states and the results are general for all topological phases. In Fig.\ref{fig8} we show two other cases with the fillings $\rho=\frac{1}{3}, 1$, and similar results are obtained.


\begin{thebibliography}{10}
\bibitem{rev1} J. E. Moore, Nature \textbf{464}, 194 (2010).

\bibitem{rev2} M. Z. Hasan and C. L. Kane, \rmp \textbf{82}, 3045
(2010).

\bibitem{rev3} X. L. Qi and S. C. Zhang, \rmp \textbf{83}, 1057
(2011).

\bibitem{rev4} S. Q. Shen, Topological Insulators (Springer, Berlin,
2012).


\bibitem{bcla0} Y.-Z. You and C. Xu, \prb {\bf 90}, 245120 (2014).
\bibitem{bcla1} X. Chen, Z.-C. Gu, and X.-G.Wen, \prb {\bf 83}, 035107 (2011).
\bibitem{bcla2} X.-G. Wen, \prb {\bf 89}, 035147 (2014).


\bibitem{haldane} F. D. M. Haldane, Phys. Lett. A {\bf 93}, 464 (1983).

\bibitem{cold1} V. Galitski, and I. B. Spielman, Nature {\bf 494}, 49 (2013).




\bibitem{tf1} L. J. Lang, X. Cai and S. Chen, \prl {\bf 108}, 220401 (2012).

\bibitem{tf2} S. Ganeshan, K. Sun and S.D. Sarma, \prl {\bf 110} 180403 (2013).

\bibitem{su1}
W. P. Su, J. R. Schrieffer and A. J. Heeger, \prb {\bf 22}, 2099 (1980).

\bibitem{tb1} S.-L. Zhu, Z.-D. Wang, Y.-H. Chan and L.-M. Duan, \prl {\bf 110}, 075303 (2013).

\bibitem{tb2} F. Grusdt, M. Honing and M. Fleischhauer, \prl {110}, 260405 (2013).

\bibitem{tb3} X. Deng and L. Santos, \pra {\bf 89}, 033632 (2014).

\bibitem{tb4} R. Barnett, \pra {\bf 88}, 063631 (2013).

\bibitem{tb5} F. Matsuda, M. Tezuka, and N. Kawakami, J. Phys. Soc. Jpn. {\bf 83}, 083707 (2014).

\bibitem{bhm1} T. Kuhner, and H. Monien, \prb {\bf 58}, 14741R, (1998).

\bibitem{bhm2} B.-L. Chen, S.-P. Kou, Y. Zhang, and S. Chen, Phys. Rev. A {\bf 81}, 053608 (2010).

\bibitem{jordan} V.G. Rousseau, D.P. Arovas, M. Rigol, F. Hebert, G.G. Batrouni and R.T. Scalettar, \prb {\bf 73}, 174516 (2006).


\bibitem{cold2} M. Atala, M. Aidelsburger, J. T. Barreiro, D. Abanin, T. Kitagawa, E. Demler and I. Bloch, Nature Phys. {\bf 9}, 795 (2013).

\bibitem{Roati} G. Roati, C. D' Errico, L. Fallani, M. Fattori, C. Fort, M.
Zaccanti, G. Modugno, M. Modugno, and M. Inguscio, Nature
(London) {\bf 453}, 895 (2008).


\bibitem{spt} H. M. Guo, and S. Chen, \prb {\bf 91}, 041402(R) (2015); C. Yang, H. M. Guo, L.-B. Fu, and S. Chen, \prb {\bf 91}, 125132 (2015).


\bibitem{berry1} R. Resta, Rev. Mod. Phys. \textbf{66}, 899 (1994).

\bibitem{berry2} D. Xiao, M.C. Chang and Q. Niu, \rmp \textbf{82},
1959 (2010).

\bibitem{zak} J. Zak, \prl {\bf 62}, 2747 (1989).

\bibitem{sandvik} Olav F. Syljuosen, and Anders W. Sandvik, \pre {\bf 66}, 046701 (2002).


\bibitem{exact} J. M. Zhang and R.X. Dong, Eur. J. Phys. {\bf 31}, 591 (2010).


\bibitem{twist1} H. M. Guo and S. Q. Shen, \prb \textbf{84}, 195107
(2011); H. M. Guo, S. Q. Shen and S. P. Feng, \prb {\bf 86}, 085124 (2012).

\bibitem{su2}
W. P. Su and J. R. Schrieffer, \prl {\bf 46}, 738 (1981).

\bibitem{exp1}
W. S. Bakr, A. Peng, M. E. Tai, R. Ma, J. Simon, J. I. Gillen, S. Folling, L. Pollet, and M. Greiner, Science {\bf 329}, 547 (2010).

\bibitem{exp2}
J. F. Sherson, C. Weitenberg, M. Endres, M. Cheneau, I. Bloch, and S. Kuhr, Nature (London) {\bf 467}, 68 (2010).

\bibitem{exp3}
C. Weitenberg, M. Endres, J. F. Sherson, M. Cheneau, P. Schausz, T. Fukuhara, I. Bloch, and S. Kuhr, Nature (London) {\bf 471}, 319 (2011).




\end{thebibliography}
\end{document}